\documentclass[%
 reprint,
 amsmath,amssymb,
 aps,
longbibliography,
]{revtex4-1}

\usepackage{graphicx}
\usepackage{grffile} 
\usepackage{subfigure}
\usepackage[marginal]{footmisc}
\usepackage{epstopdf}
\usepackage{amssymb}
\usepackage{mathrsfs}
\usepackage{amsmath,mathtools,amsfonts,bm,graphicx}
\usepackage{upgreek}
\usepackage{color}
\usepackage{etoolbox}
\usepackage[colorlinks=true,linkcolor=blue,urlcolor=blue,citecolor=blue]{hyperref} 
\usepackage{multirow}
\usepackage{mathtools}
\usepackage[normalem]{ulem}
\usepackage[colorlinks=true,linkcolor=blue,urlcolor=blue,citecolor=blue]{hyperref}

\newcommand{\fig}[1]{Fig.~\ref{#1}}
\newcommand{\Fig}[1]{Figure~\ref{#1}}
\newcommand{\eq}[1]{Eq.~(\ref{#1})}
\newcommand{\eqr}[2]{Eqs.~(\ref{#1})-(\ref{#2})}
\newcommand{\eqs}[2]{Eqs.~(\ref{#1}) and (\ref{#2})}

\newcommand{\blue}[1]{{\textcolor{blue}{#1}}}

\renewcommand{\Ref}[1]{Ref. [\onlinecite{#1}]}

\newcommand{\roundbk}[1]{\left( #1 \right)}

\newcommand{\squarebk}[1]{\left[ #1 \right]}

\newcommand{\abs}[1]{\lvert #1 \rvert}

\newcommand{\DEmax}{{\Delta E^{max}}}
\newcommand{\DEabs}{{\abs{\Delta E}}}
\newcommand{\qcool}{\nu_{cool}}
\newcommand{\f}{\Phi}
\newcommand{\Q}{q}
\newcommand{\pvoid}{p^{void}_i}
\newcommand{\Ti}{T_{I}}

\begin{document}

\title{Emergence of two-level systems in glass formers: a kinetic Monte Carlo study}

\author{Xin-Yuan Gao$^1$}
\author{Hai-Yao Deng$^2$}
\author{Chun-Shing Lee$^1$}
\author{J. Q. You$^3$}

\author{Chi-Hang Lam$^1$}
\email[Email: ]{C.H.Lam@polyu.edu.hk}
\address{$^1$Department of Applied Physics, Hong Kong Polytechnic University, Hong Kong, China \\
$^2$School of Physics and Astronomy, Cardiff University, 5 The Parade, Cardiff CF24 3AA, Wales, UK\\
$^3$Department of Physics, Zhejiang University, Hangzhou 310027, China
}

\date{\today}

\begin{abstract}
Using a distinguishable-particle lattice model based on void-induced dynamics, we successfully reproduce the well-known linear relation between heat capacity and temperature at very low temperatures. The heat capacity is dominated by two-level systems formed due to the strong localization of voids to two neighboring sites, and can be exactly calculated in the limit of ultrastable glasses. Similar but weaker localization at higher temperatures accounts for the glass transition. The result supports the conventional two-level tunneling picture by revealing how two-level systems emerge from random particle interactions, which also cause the glass transition. Our approach provides a unified framework for relating microscopic dynamics of glasses at room and cryogenic temperatures.
\end{abstract}

\maketitle
\section{Introduction}
\label{sec:introduction}
Most liquids can be quenched into the glassy state by undergoing a  glass transition, a phenomenon actively studied for decades \cite{stillinger2013review,arceri2020}. When further cooled below $\sim$1K, it was found by Zeller and Pohl that the heat capacity of glasses is proportional to the temperature $T$, well exceeding Debye's $T^3$ relation based on acoustic phonons \cite{pohl1971}. Anderson \textit{et al}~\cite{anderson1972} and Phillips~\cite{phillips1972} simultaneously proposed that the heat capacity is dominated at low $T$ by two-level systems (TLS). Their theory has successfully explained a plethora of low-$T$ thermal and acoustic properties of glasses~\cite{phillips1987}. Nevertheless, the microscopic nature of TLS and their possible universal properties remain controversial~\cite{leggett2013,queen2013,ramos2014,ramos2020,carruzzo2020}. Recently, TLS in glasses have attracted additional interest due to their strong relevance to noise in quantum computing devices \cite{muller2019}.

Numerous glasses \cite{ramos2020} exhibit the characteristic heat capacity found in \Ref{pohl1971}. Therefore, TLS is likely an intrinsic component in glasses and should be relevant to the glass transition and glassy dynamics in general. Yet, TLS at present plays little role in major theories of glass transition \cite{stillinger2013review,arceri2020}. Concerning particle simulations, both molecular dynamics (MD) simulations \cite{kob1995} and lattice models \cite{garrahan2011review} can reproduce many features of glasses. Identification of TLS in MD systems has been reported \cite{damart2018,khomenko2020}. However, according to Refs. \cite{anderson1972,phillips1972}, the characteristic low-$T$ heat capacity depends not only on the existence of TLS, but also that they must be sufficiently isolated from each other. The latter condition has not been fully explored in any particle simulation and, more importantly, the hallmark low-$T$ heat capacity has not been explicitly reproduced.
As MD simulations become computationally challenging at low $T$ due to the slow dynamics, accessing the heat capacity directly can be difficult.  
Neither has this been achieved in conventional lattice models, despite their better computational efficiencies \cite{garrahan2011review}.
On the other hand, heat capacity linear in $T$ has also been shown to be explainable with diffusive vibrational modes \cite{baggioli2019hydrodynamics} and observed in a random network model \cite{baggioli2019} that apparently exhibits no TLS.

In this work, we successfully reproduce the characteristic low-$T$ heat capacity of glasses using a recently proposed distinguishable particle lattice model (DPLM), which has already been shown to exhibit typical glass transition \cite{zhang2017}. The heat capacity is shown to be dominated by TLS, which naturally emerge from increasingly strong particle localization as $T$ decreases. We demonstrate that the same localization effects are responsible for the glass transition at higher $T$.

At $T \lesssim 10K$, the heat capacity of many glasses follows
$c_1 T + (c_D + c_3) T^3$ ~\cite{pohl1971,ramos2020}. The linear term $c_1T$ dominates at $T\lesssim 1K$ and is explained by the TLS theory \cite{anderson1972,phillips1972}. The Debye contribution $c_D T^3$ can be independently determined from acoustic properties. Results in general support the existence of an extra $c_3 T^3$ term, which can be approximately accounted for using soft-potential models \cite{karpov1983,buchenau1991}.
Being a lattice model, the DPLM does not accommodate vibrations, leading to $c_D=0$. We will show below that under a wide range of conditions, the specific heat capacity $C_v$ of the DPLM at low $T$ follows
\begin{equation}
  \label{Cv}
  C_v = c_1 T + c_3 T^3,
\end{equation}
consistent with experiments.

The DPLM has been shown to exhibits typical glassy behaviors such as a pronounced plateau in the mean-squared displacement of particles~\cite{zhang2017,deng2019} and stretched exponential relaxation in the self-intermediate scattering function~\cite{zhang2017}. It has recently
afforded an explanation of the decades-old Kovac's expansion gap paradox~\cite{lulli2020}, reproduced Kovacs memory effect~\cite{lulli2021}, suggested simple connections among glass fragility, entropy and particle pair-interactions~\cite{lee2020} and demonstrated heat-capacity overshoot~\cite{lee2021} . The present demonstration of characteristic low-$T$ thermal properties in the same model thus establishes a unique framework to relate the TLS theory to the rich dynamical behaviors of glasses at higher $T$. %
\section{Model}
\label{sec:model}
We adopt basically the DPLM defined in \Ref{lee2020}. 
It is a two-dimensional lattice model with $N$ distinguishable particles. Each particle has its own type, and can move on a square lattice of size $L^2$. A vacant lattice point is deemed occupied by a void so that the void density is $\phi_v={1 -N}/{L^2}$. The system has a total energy
\begin{equation}
E=\sum_{{<i,j>}'}V_{s_is_j}
\label{E}
\end{equation}
where the sum is only applied to occupied adjacent sites $i$ and $j$. There are thus only nearest neighboring interactions between particles in the model. The index $s_i = 1,2, \dots , N$ denotes which particle is at sites $i$. 
Each interaction $V_{kl}$ between particle $k$ and $l$ is sampled randomly from a distribution $g(V)$.
The dynamics is furnished by the Metropolis rule satisfying detailed balance: each particle can hop to an empty adjacent site (i.e. a void) at a rate
\begin{equation}
w(\Delta E) = 
\begin{cases}
w_0 e^{- \Delta E / k_BT} & \text{for} ~ \Delta E>0, \\
w_0  & \text{for} ~ \Delta E\leq 0, \label{w}
\end{cases}
\end{equation}
where $k_B=1$, $w_0=10^{6}$, and $\Delta E$ is the change of the system energy $E$ due to the hop. Notice that the particle indices $s_i$ and $s_j$ are implicitly time dependent, since particles move around.%
\section{Specific heat measurement}
\label{sec:specific_heat_measurement}
In our main simulations, we consider for simplicity an interaction distribution $g(V)$ uniform over $[V_0,V_0+\Delta V]$, where $\Delta V=1$. We put $V_0 =0$, corresponding to purely repulsive interactions which suppress void aggregation even at low $T$. \blue A general form of $g(V)$ should give qualitatively similar results. As will be discussed below, a uniform $g(V)$ does not gives rise to, and should not be confused with, a flat TLS energy distribution. The latter  is a commonly used simplification but is again non-essential for arriving at the experimental low-$T$ heat capacity \cite{anderson1972,phillips1972}.  

\begin{figure}[tpb]
\centering
\includegraphics[width=0.45\textwidth]{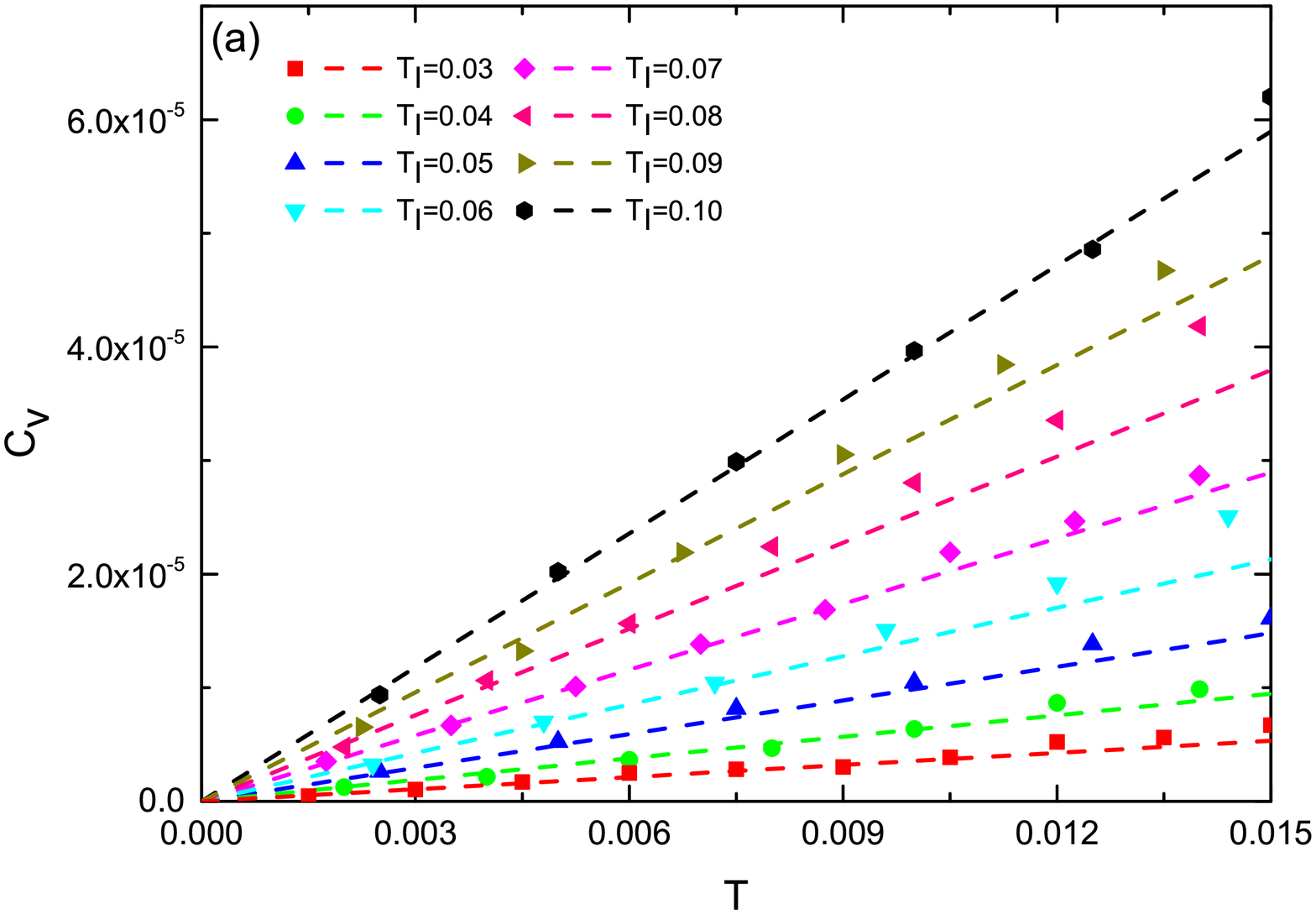}
\includegraphics[width=0.45\textwidth]{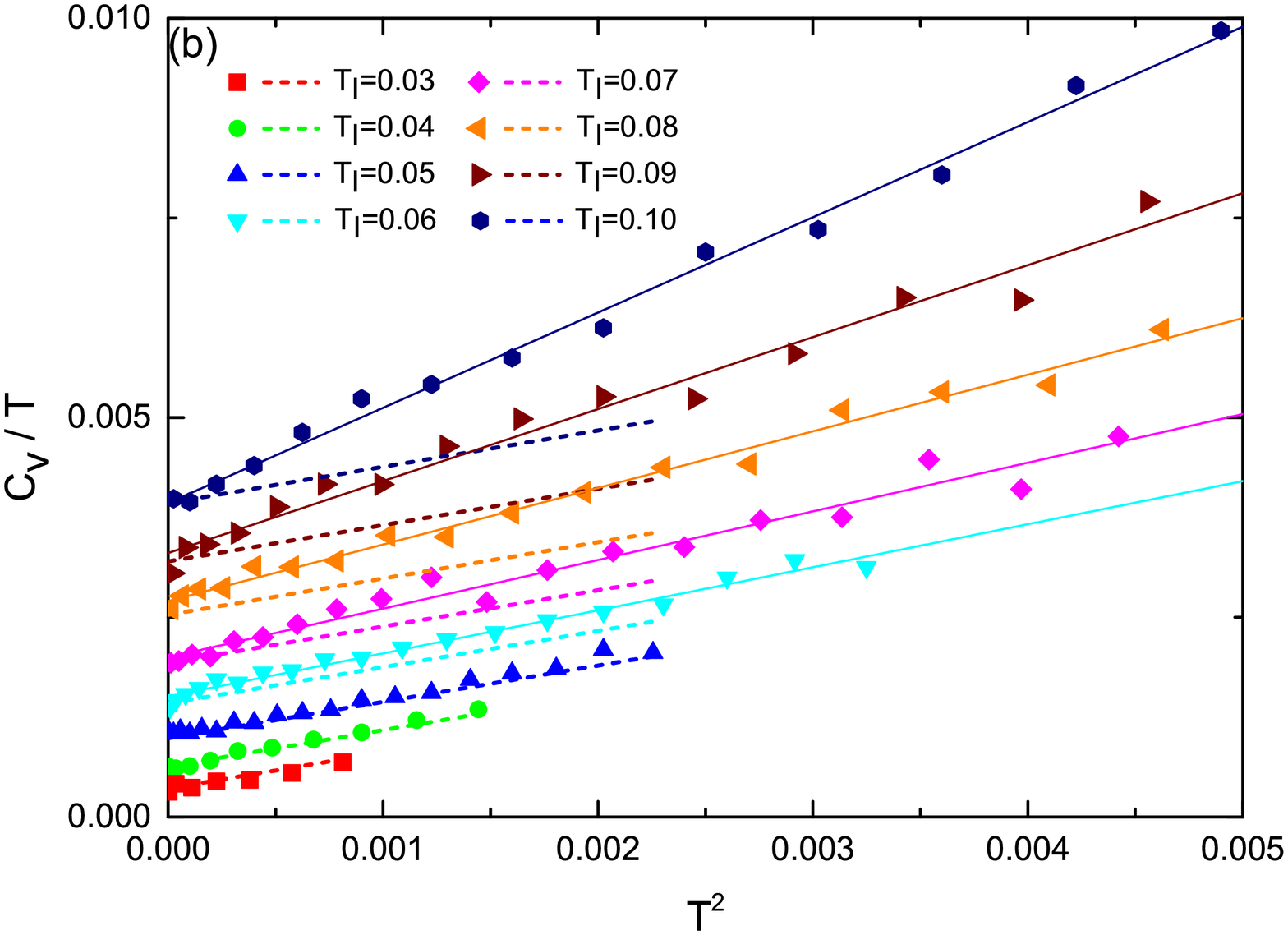}
\caption{(a) Specific heat capacity $C_v$ against temperature $T$ for various initial temperature $\Ti$. (b) Plot of $C_v/T$ against $T^2$ using the same data as in (a) but over a wider range of $T \le \Ti$. (a) Dashed lines and (b) short dashed lines show theoretical values from \eqs{Cv}{c1c2}, which are accurate for $\Ti \alt 0.05$. For $\Ti \agt 0.06$, \eq{Cv} remains valid as shown by their linear fits (solid lines). 
 }
\label{CvT}
\end{figure}

We initialize equilibrium systems on a $200\times 200$ lattice with a void density $\phi_v=0.005$ at  temperature $\Ti$ via direct construction~\cite{zhang2017}. Kinetic Monte Carlo simulations are then  performed with $T$ decreasing from $\Ti$ towards $0$ at a cooling rate 
$\qcool=10^{-4}$. 
We continuously measure the system energy $E$ defined in \eq{E} so as to calculate $C_v=N^{-1}~ dE/dT$. 
The glass transition temperature in our system is found to be $T_g \simeq 0.15$, which has been defined as the temperature at which the particle diffusion coefficient $D$ falls to a small reference value $D_{r} \equiv 0.1$ \cite{lee2020}. 
We first consider low initial equilibrium temperatures $\Ti \ll T_g$, leading to ultrastable glasses \cite{swallen2007,zhao2013} with a low fictive temperature close to $\Ti$. 
Simulation results on $C_v$ are plotted in \fig{CvT}(a). We observe that the DPLM successfully reproduce the linear relation between $C_v$ and $T$, i.e. \eq{Cv} in the low $T$ limit. 
Moreover, we find that $c_1$, which equals the slope, decreases with $\Ti$. 
The reduction of $c_1$ shows a depletion of TLS, fully consistent with suggestions based on experiments \cite{queen2013,ramos2014}. 

After confirming the $c_1T$ term, we now examine the full expression in \eq{Cv}. \Fig{CvT}(b) plots $C_v/T$ against $T^2$. The reasonable linear relations observed in all cases verify \eq{Cv} with $c_3>0$. 
Similar to experimental results \cite{pohl1971}, the absence of any second order term, i.e. $c_2 T^2$, is evident.
Nevertheless, the presence of the $c_3T^3$ term is, at first sight, surprising, since similar nonlinear terms such as a $T^5$ term has been suggested to be accounted for by the soft-potential model concerning anharmonic vibrations \cite{karpov1983,buchenau1991}. It is somewhat not expected for a lattice model. This will be discussed later.

\begin{figure}[tpb]
\centering
\includegraphics[width=0.5\textwidth]{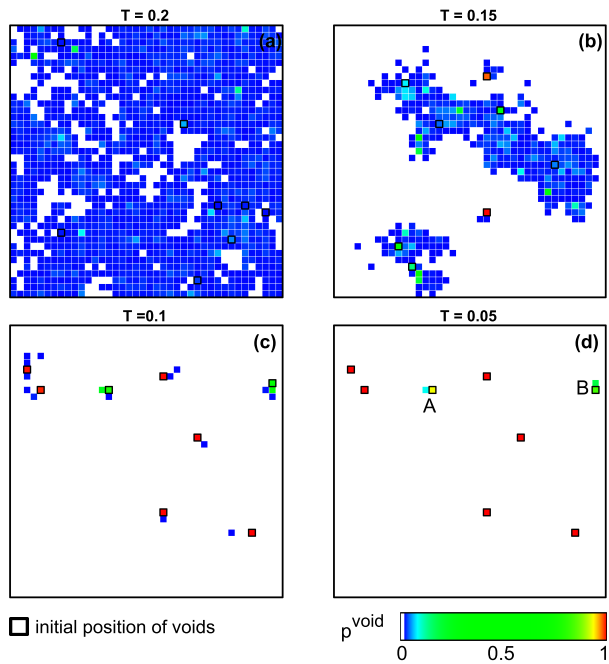}
\caption{ Spatial profile showing occupation probability $\pvoid$ of voids at site $i$ during cooling when temperature $T$ reaches 0.2 (a), 0.15 (b), 0.1 (c)  and 0.05 (d). In each case, $\pvoid$ is measured over a period during which $10^7$ particle hops have occurred. Hops appear fewer at lower $T$ because of increasingly severe back-and-forth motions. Sites at which no void is detected are shaded white. Initial void positions at each period are marked by black squares. A pair of two-level systems A and B have emerged in (d).
}
\label{pvoid}
\end{figure}
\section{Particle dynamics}
\label{sec:real_space_image}
\newcommand{\Pret}{P_{ret}}
\newcommand{\Pretv}{P^{void}_{ret}}
A close examination of the particle dynamics shows that stronger particle localization at low $T$ accounts for both the glass transition and the emergence of TLS. In the DPLM, 
particle movements are induced by voids, a mechanism supported by recent colloidal experiments \cite{yip2020}.
Since a particle hop can be equivalently considered as the opposite hop of a void, we describe the dynamics of particles and voids interchangeably.
\Fig{pvoid} shows spatial profiles of the void occupation probability $\pvoid$ at site $i$ on a $40\times 40$ lattice at different stages of cooling.
To enable a meaningful comparison, $\pvoid$ in each case is measured over a period of time during which $10^7$ particle hops have occurred.

For $T = 0.2 \gg T_g$ corresponding to the non-glassy liquid phase, we observe  that voids diffuse quite freely. 
Thermal excitations dominate over random particle interactions. 
When cooled to $T= 0.15\simeq T_g$, $\pvoid$ is much more heterogeneous, with   highly preferential sites of locating voids.
Such void localization is caused by the random particle interactions. It leads to significant dynamic slowdown and thus the glass transition as characterized in Refs. \cite{zhang2017,lee2020} and will be further quantified below.
As the system is further cooled to $T = 0.1 \ll T_g$, most voids are persistently trapped to within a few sites. Some of them are even completely frozen. The system can no longer fully relax within practical simulation time, implying the glass phase. At $T = 0.05$, the strong localization completely freezes most voids. More importantly, a small number of voids are trapped between only two sites, forming TLS. 

\begin{figure}[t]
\centering
\includegraphics[width=0.22\textwidth]{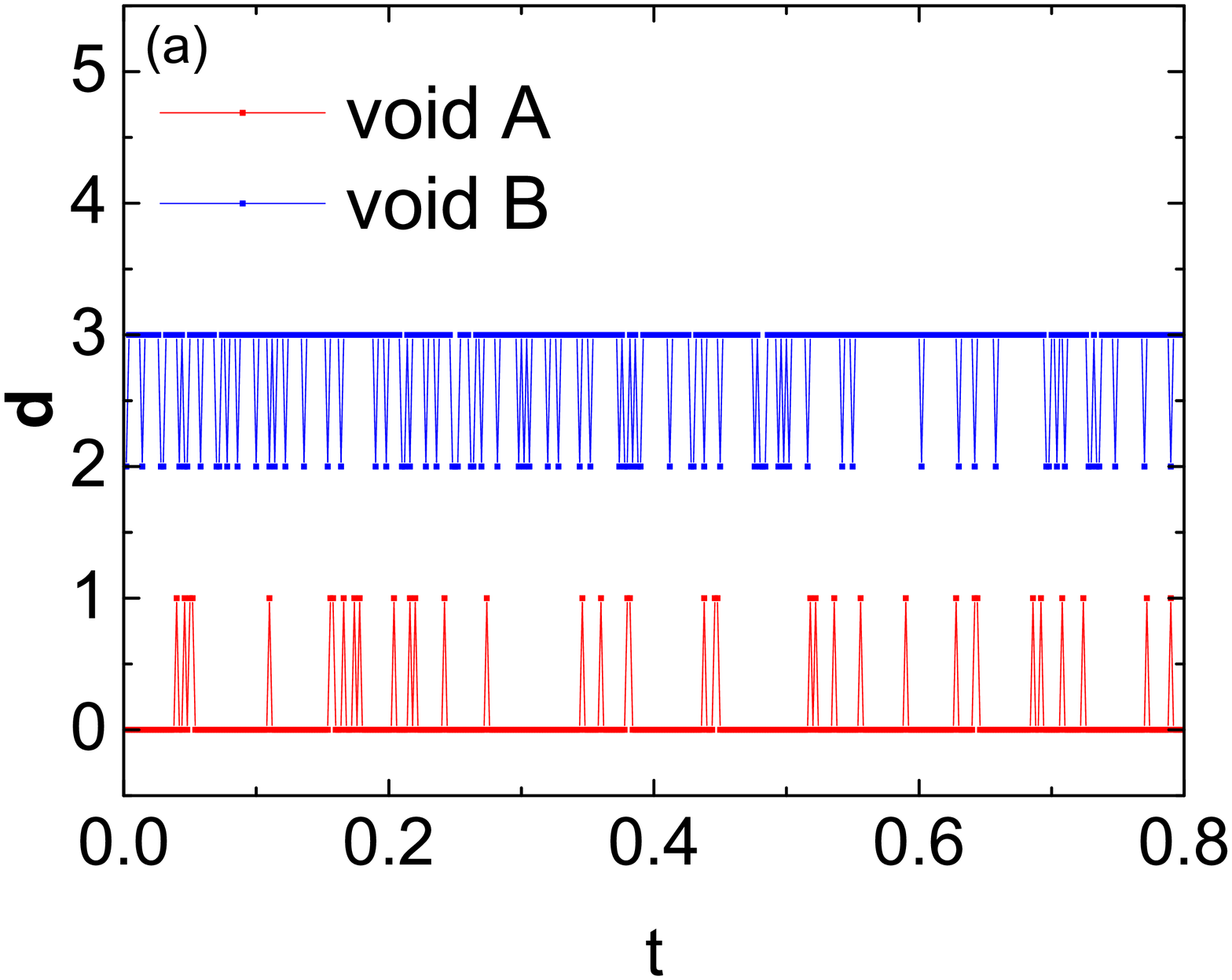}
\includegraphics[width=0.22\textwidth]{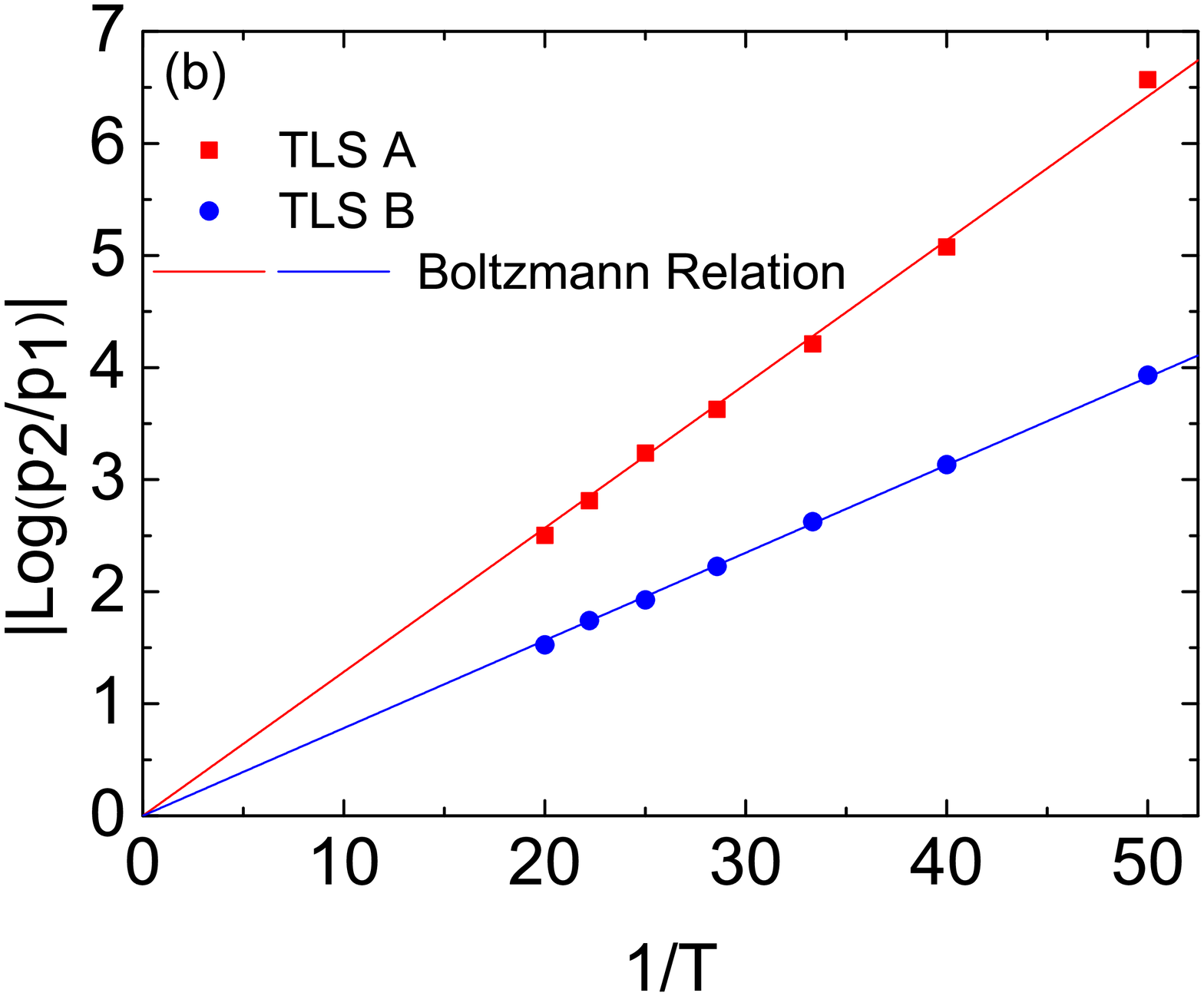}
\caption{(a) Plot of void displacement $d$ against time $t$ of the two TLSs in \fig{pvoid} at $T=0.05$. Results for TLS B are shifted upward for clarity. 
(b) Plot of $\abs{\log(p_2/p_1)}$ against $1/T$ for the same TLSs in (a),  where $p_1$ and $p_2$ are the measured probabilities of the two levels in a TLS. Solid lines are fits to the Boltzmann relation.
}
\label{voiddisp}
\end{figure}
\begin{figure}[t]
\centering
\includegraphics[width=0.22\textwidth]{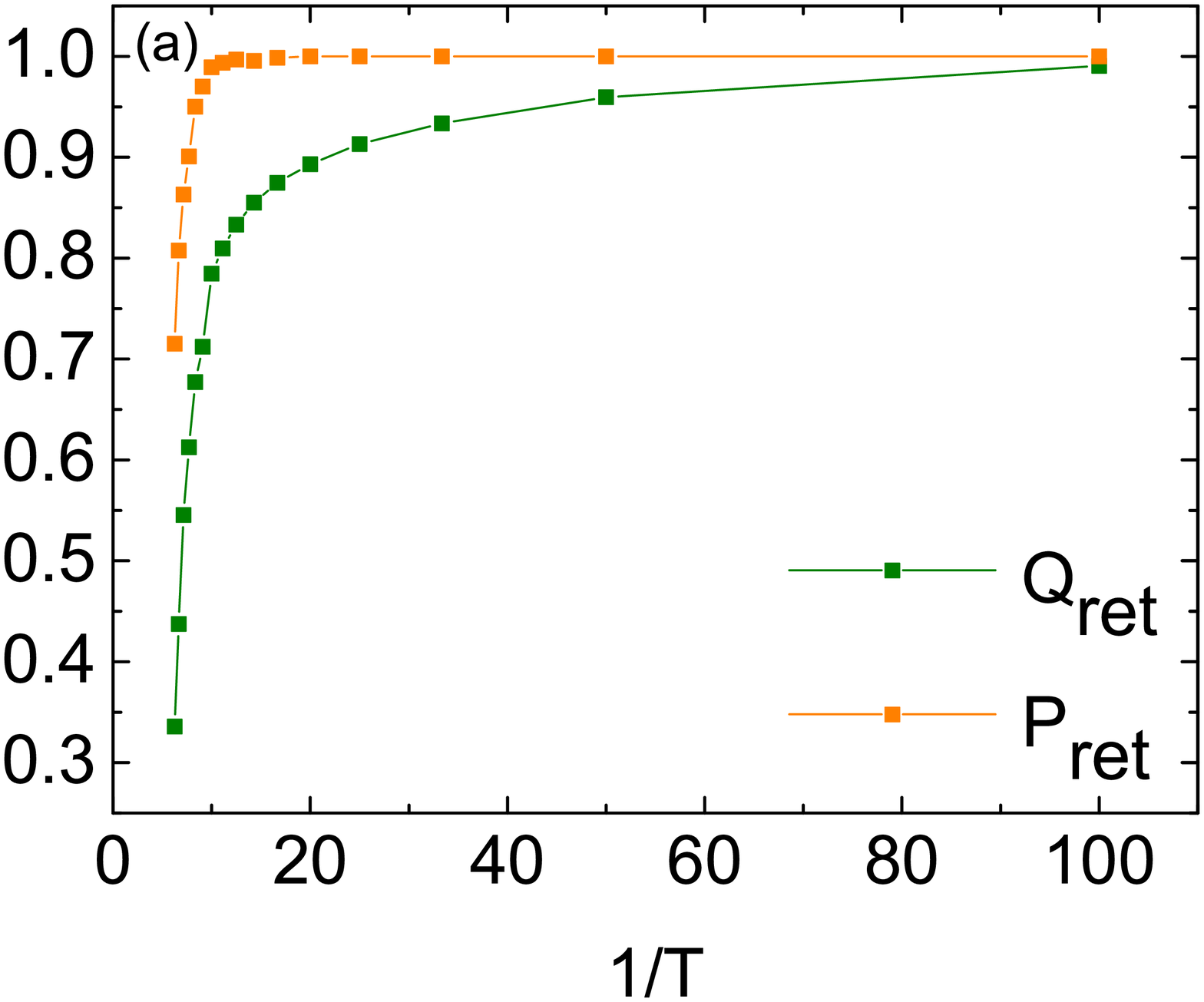}
\includegraphics[width=0.24\textwidth]{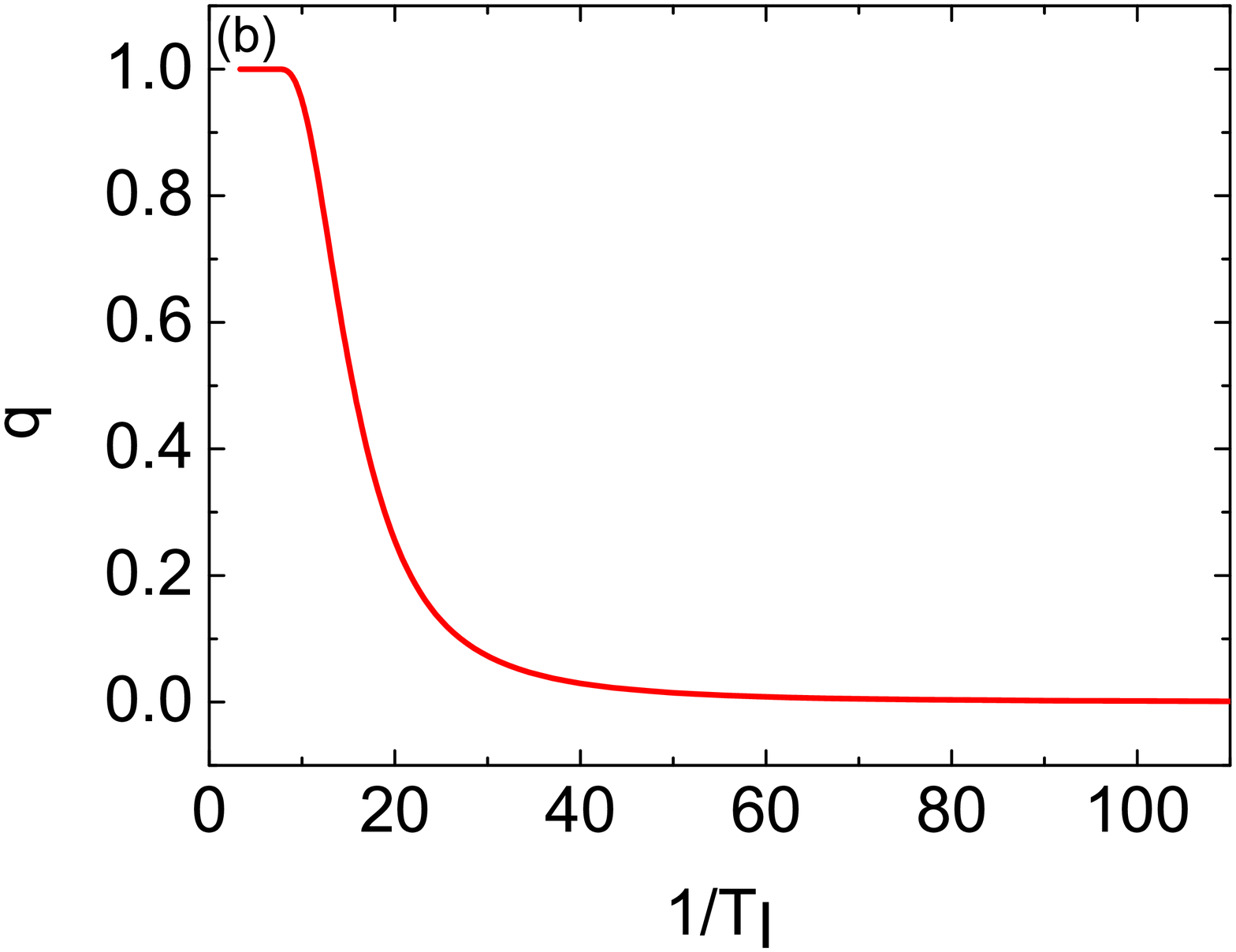}
\caption{(a) Particle and void return probabilities, $P_{ret}$ and $Q_{ret}$ against  $1/T$. (b) Probability $q$ that a particle hop is energetically possible against $1/\Ti$. 
}
\label{Pret}
\end{figure}%
The TLS in our system exemplified in \fig{pvoid} are isolated and noninteracting, due to the strong localization and the small void density $\phi_v=0.005$ used. We emphasize that noninteracting or weakly interacting TLS are essential to account for the experimental $C_v$ \cite{anderson1972,phillips1972}. At $T \alt 0.05$, further system relaxation is limited to TLS transitions, while TLS movements, restructuring and other relaxations are all negligible.
\Fig{voiddisp}(a) shows displacement-time graphs of the voids constituting the two TLS in \fig{pvoid}. The bistability is evident, with each level corresponding to the void at one of the two energetically possible sites. The occupation probabilities $p_1$ and $p_2$ of the initial and the hopped levels are asymmetrical in general and depend on the energy difference $\Delta E$. \Fig{voiddisp}(b) plots $\left| \log  p_2 / p_1 \right |$ against $1/T$. The nice linearity obtained verifies the equilibrium relation $p_2 / p_1 = \exp( -\Delta E /k_B T)$. 
Hence, TLS form equilibrium subsystems, in sharp contrast to the whole system which is out of equilibrium.

To further quantify in a unified manner how localization induces both the glass transition and TLS, we study the hopping return probabilities $P_{ret}$ and $Q_{ret}$ of particles and voids respectively. After a particle has hopped, the return probability $P_{ret}$  is defined as the probability that the next hop by the particle reverse its previous hop and return it to the original position \cite{lam2017,zhang2017,lee2020,yip2020}. Here, we define $Q_{ret}$ analogously for voids. 
\Fig{Pret} plots $P_{ret}$ and $Q_{ret}$ against $1/T$ during cooling. Results are measured from snapshots of system configurations. They provide lower bounds of the probabilities, since some rapid back-and-forth motions in between consecutive snapshots may not be registered \cite{lam2017}. At high $T$, both $P_{ret}$ and $Q_{ret}$ are relatively small as dynamics are closer to random walks. During cooling, they decrease monotonically and smoothly. At $T =0.15 \simeq T_g$, we get $P_{ret}\simeq 0.8$. This implies a strong back-and-forth nature of the particle hops
\cite{vollmayr2004,lam2017}, which is a main contributor to the dramatically slowed down dynamics at the glass transition \cite{lee2020}.
At $T=0.10$, $P_{ret} \simeq 0.99$. Since nearly all hopping motions are reversed, particle dynamics are basically arrested, evidencing that the glass transition has already occurred and the system is deeply in the glass phase.
At $T=0.01$, $Q_{ret} \simeq 0.99$, showing that nearly all dynamics are TLS transitions. 
Note that although particle hops are induced by voids, $P_{ret} > Q_{ret}$ at all $T$. 
This can be understood by noting, for example, that two consecutive non-returning hops by a single void involve single-hops by two different particles, resulting at distinct statistics for particles and voids.

\section{Emergence of two-level systems}
\label{sec:Emergence of two-level systems}
A unique feature of the DPLM is its exact equilibrium properties \cite{zhang2017} which have been extensively verified numerically \cite{zhang2017,lulli2020,lee2020}. 
This allows us to analytically deduce the emergence of TLS as follows.
Let $\Delta E$ be the system energy change due to a hop attempt of a particle into a nearest neighboring void. The probability distribution $P(\Delta E)$ can be computed for equilibrium systems \cite{lam2018}, but in general depends non-trivially on the thermal history for non-equilibrium systems.
The probability $q$ that a hop 
is energetically possible can be approximated by
\begin{equation}
\Q=  \int^{\DEmax}_{-\infty}P(\Delta E)\mathrm{d}\Delta E
\label{q}
\end{equation}
where $\DEmax$ is the maximum energy cost for a hop attempt to be considered energetically possible. During cooling, temperature is close to $T$ for a duration $\tau$ which, as an order of magnitude estimation, is given by $\tau \sim {0.1 ~T/\qcool}$. For at least one hop to occur during $\tau$, the hopping rate must satisfy $w  \agt 1/\tau$, which gives
\begin{equation}
  \label{DEmax}
\DEmax \simeq \ln(0.1 ~ w_0 T  / \qcool) k_B T.
\end{equation}
after using \eq{w}.
For systems equilibrium at  $\Ti$,  
 $\Q$ is calculated using \eqr{q}{DEmax} and exact expressions of $P(\Delta E)$ from \Ref{lam2018} and results are plotted in \fig{Pret}(b).

At small $\Ti$, we observe that $\Q$ converges towards 0, e.g. $\Q \simeq 0.05$ at $\Ti=0.03$.  Voids then have vanishingly few energetically possible hopping pathways. Most voids are thus frozen. Some voids possess one energetically possible hop with a probability  ${{\sim}z \Q}$, where $z=4$ is the lattice coordination number.
Each then forms a TLS leading to a TLS density
$\phi_{TLS} \simeq z \Q \phi_v  \label{phiTLS}$.
If a void is allowed multiple possible hops, a multi-level system with three or more levels results. This however occurs at a probability of order $\Q^2$ or smaller and are negligibly few compared with TLS. 

The above analysis is directly applicable to $T \le \Ti \ll T_g$ corresponding to  ultrastable glasses. 
Most glasses are however less stable with a fictive temperature around $T_g$. The above picture is still qualitatively applicable because once cooled to $T \ll T_g$, most dynamics are frozen, as can be observed from \fig{pvoid}. Hence, $\Q$ should similarly approach 0.
Nevertheless, due to local relaxations predominantly in the vicinity of voids, the system is overall  non-equilibrium so that  $P(\Delta E)$, $q$ and $\phi_{TLS}$ cannot be calculated analytically.

We  begin our derivation of \eq{Cv} by assuming ${T \ll T_g}$.
As explained above, most voids are completely frozen and have null contribution to $C_v$. 
Voids forming multi-level systems are on the other hand few and can be neglected. 
Therefore, we only need to consider the TLS which dominate $C_v$.
Since TLS are at equilibrium as shown in \fig{voiddisp}(b), straightforward algebra gives
\begin{equation}
C_v = z \phi_{v}  \int^\DEmax_{-\infty} d\Delta E ~ {P}(\Delta E) \f(\Delta E), \label{Cv2}
\end{equation}
where 
\begin{equation}
\f(\Delta E) = \frac{1}{4k_BT^2} \Delta E^2  \text{sech}^2\left(\frac{\Delta E}{2k_BT}\right)
\end{equation}
is the heat capacity of a TLS~\cite{phillips1972}. Note that $\f(\Delta E)$ is an even function peaked sharply at $\Delta E \approx \pm 2.35 k_BT$.
This physically represents that TLS with large energy splits contribute little to $C_v$. The upper integration limit can thus be approximated as infinity, giving
\begin{equation}
C_v = z \phi_{v}  \int^\infty_{0} d \DEabs ~ \tilde{P}(\DEabs) \f(\DEabs).
\label{Cvtheory}
\end{equation}
Here, $\DEabs$  is the TLS energy split with a distribution
\begin{equation}
\tilde{P}(\DEabs) = P(\Delta E) + P(-\Delta E).
\label{Ptilde}
\end{equation}
This expression highlights the equivalent contributions to $C_v$ by hops with positive and neglect energy changes. 
Conventionally,
$\tilde{P}(\DEabs)$ is assumed a constant for simplicity \cite{anderson1972,phillips1972}.  Instead, we expand $\tilde{P}(\DEabs)$ about $\DEabs=0$, keeping only the first two non-zero terms.
After some algebra, \eq{Cvtheory} reduces to \eq{Cv} with
\begin{equation}
c_1 = \frac{\pi^2z\phi_v k^2_B P(0)}{3}, ~~~~~
c_3 = \frac{7\pi^4z\phi_v k^4_B P^{''}(0)}{15}. 
\label{c}
\end{equation}
All even terms, e.g. $c_2 T^2$, vanish exactly since $\tilde{P}(\DEabs)$ is even.

For ultrastable glasses with $T \le \Ti \ll T_g$, we can calculate $c_1$ and $c_3$ using exact expressions of $P(\Delta E)$ \cite{lam2018}. In particular, for the uniform interaction distribution $g(V)$ used in our main simulations and $z=4$, we get
\begin{equation}
c_1 = \frac{8\pi^2 \phi_v k^4_B \Ti^2}{ \Delta V^3},  ~~~~~~ c_3 = \frac{14\pi^4\phi_vk^4_B}{15\Delta V^3},
\label{c1c2}
\end{equation}
which is exact for $\Ti\rightarrow 0$.
Figure \ref{CvT} plots $C_v$ from \eqs{Cv}{c1c2}. We observed an excellent agreement with simulations for $c_1$ and $c_3$ at $\Ti \alt 0.10$ and 0.05 respectively.
For less stable glasses with higher $\Ti$, discrepancies from \eq{c1c2} occur. This is because local relaxation results in deviation from the exact form of $P(\Delta E)$ at $\Ti$ used in our calculations.  Nevertheless, \eq{Cv} remains valid. See appendix \ref{sec:ana_cal} for more detailed calculations.

\section{Comparison with glycerol}
\label{sec:comparison_with_glycerol}

The DPLM is a microscopic model allowing quantitative comparisons with real materials  \cite{lee2020}.
\begin{figure}[tpb]
\centering
\includegraphics[width=0.45\textwidth]{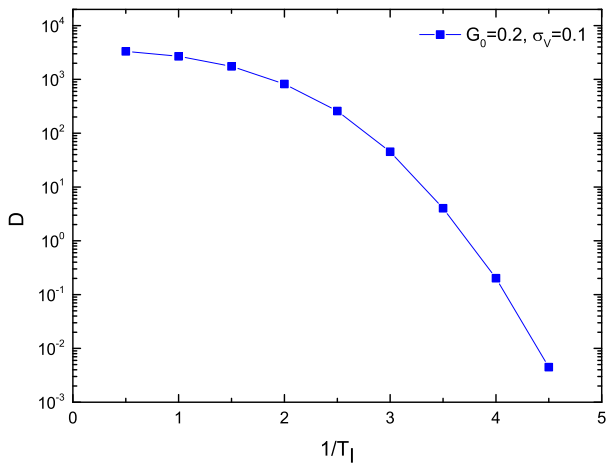}
\caption{Arrhenius plot of diffusion coefficient $D$ against $1/T$ for particle interaction distribution $g(V)$ in a Fermi-plus-Gaussian form for $G_0=0.2$ and $\sigma_V=0.1$ indicating a moderately fragile glass. We take $\phi_v=0.01$. 
}
\label{Fig:D}
\end{figure}
Up to now, we have been using dimensionless units. A quantitative comparison with real materials requires using physical units, which will be adopted in this section. First, we follow \Ref{lee2020} to match the kinetic fragility of a specific material. 
To generate glasses of various fragilities, an energetic parameter $G_0 \in [0, 1]$ is introduced in the interaction energy distribution $g(V)$ by generalized it to a uniform-plus-delta functional form
\begin{equation}
g(V)=G_0+(1-G_0)\delta (V-\Delta V) 
\label{si:gV1}
\end{equation}
where $\delta$ denotes Dirac's delta function. To suppress void aggregation, we take in this work  $0\leq V \leq \Delta V = 1$ corresponding to fully repulsive interactions.
When $G_0 = 1$, \eq{si:gV1} reduces to the uniform $g(V)$ adopted in our main simulations. For small but finite $G_0$, we get fragile glasses \cite{lee2020}.

At small $G_0$, the physical significance of $g(V)$ in \eq{si:gV1} is that it includes  a high-entropy high-energy (delta) component and a low-entropy low-energy (uniform) component. It was shown that replacing the delta function by a narrow Gaussian function gives similar results \cite{lee2020}. To eliminate non-analyticities  which adversely impact our calculations, we further generalize \eq{si:gV1} to a Fermi-plus-Gaussian form:
\begin{equation}
g(V)= G_0 f_{\tt Fermi}(V) + (1-G_0) f_{\tt Gau}(V) 
\label{si:gV2}
\end{equation}
where
\begin{eqnarray}
f_{\tt Fermi}(V) &=&
\squarebk{1+\exp\roundbk{\frac{V-\Delta V}{\sigma_V}}}^{-1}\\
f_{\tt Gau}(V) &=&
\frac{1}{\sqrt{2\pi}\sigma}\exp \squarebk{-\frac{(V-\Delta V)^2}{2\sigma_V^2}}
\label{si:S3e}
\end{eqnarray}
for $V\ge 0$. Here, $\sigma_V$ denotes the width of both the Gaussian and the drop in the Fermi function. When $\sigma_V\rightarrow 0$, \eq{si:gV1} is recovered. 

We have performed DPLM simulations using $g(V)$ in \eq{si:gV2} with $G_0=0.2$ and a cooling rate $\nu_{cool}=10^{-2}$. 
\fig{Fig:D} shows the measured diffusion coefficient $D$ against $1/T$. The kinetic fragility measured based on a reference diffusion coefficient $D_r = 0.1$  is $m_k=13$. Extrapolating to $D_r=10^{-14}$ following \Ref{lee2020}, we get $m_k=50$ which is comparable to the value 53 for glycerol \cite{angell1997}. 
As explained in \Ref{lee2020}, the bi-component $g(V)$ we adopt is closely related to the bond excitation model of Moynihan and Angell \cite{angell2000}. The bond excitation model uses two parameters to describe thermodynamic properties of different materials: the entropy difference $\Delta S_0$ and the enthalpy difference $\Delta H_0$ between an unexcited and an excited state. They correspond to the Fermi (uniform) and Gaussian (delta) part of the bi-component $g(V)$, respectively. We can calculate $\Delta S_0$ and $\Delta H_0$ of the two components in DPLM following \Ref{lee2020}. For $G_0=0.2, T_g=0.24$, one can find that $\Delta S_0/k_B\approx \ln [(1-G_0)/G_0] =1.39$ and $\Delta H_0/k_BT_g\approx (1-T_g)/T_g =3.17$. This is in agreement with a fit to the experimental thermodynamic data of glycerol using the bond excitation model, which gives $\Delta S_0/k_B=1.65$ and $ \Delta H_0/k_BT_g=3.84$ \cite{angell2000}.
\begin{figure}[tpb]
\centering
\includegraphics[width=0.45\textwidth]{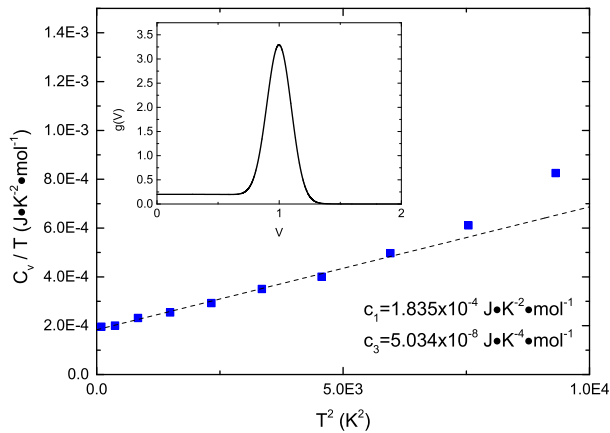}
\caption{Plot of $C_v/T$ against $T^2$ during cooling. Parameters are $G_0=0.2$, $\sigma_V=0.1$, $\phi_v=0.031$, and $k_B=1$ so that $T_g=0.24$ in dimensionless units. Results are converted to physical units appropriate for glycerol by taking $T_g=193K$ and $k_B=8.314J/(K\cdot mol)$. Inset: particle interaction distribution $g(V)$ in a Fermi-plus-Gaussian form.
}
\label{Fig:Se}
\end{figure}
The two parameters $\phi_v$ and $\sigma_V$ have weaker impacts on the fragility. In general, $c_1$ increases with $\phi_v$ while the dependence on $\sigma_V$ is non-monotonic. We take $\sigma_V = 0.1$, resulting at the distribution $g(V)$ shown in the inset of \fig{Fig:Se}. We then find that taking $\phi_v = 0.031$ provides a reasonable value of $c_1$. In dimensionless unit with $k_B=1$, we get $c_1=0.0178$ and $T_g=0.24$. To convert to physical units, we note that $T_g=193K$ for glycerol \cite{angell1997} and $k_B=8.314$ J/(K$\cdot$mol). This gives $c_1 = 1.835\times 10^{-4}$ J/K$^2$ mol. It matches  the experimental value of $c_1=1.84\times 10^{-4}$ J/K$^2$ mol \cite{stephens1973}. However, we get from simulations $c_3=5.034\times 10^{-8}$ J/K$^4$ mol, which is a few orders smaller than $c_3=1.01\times 10^{-3}$ J/K$^4$ mol from experiments \cite{stephens1973}. Therefore, using realistic model parameters, our model provides a possible explanation of $c_1$, while $c_3$ has to be approximately accounted for using other mechanisms such as lattice vibrations considered by the soft-potential model \cite{karpov1983,buchenau1991}. %
\section{Discussions}
\subsection{Comparison with standard TLS picture}

Anderson \textit{et al}~\cite{anderson1972} and Phillips~\cite{phillips1972} proposed the standard TLS model which explains the characteristic low-$T$ heat capacity, heat conductivity, phonon echoes properties of glasses ~\cite{phillips1987}. While some properties such as heat capacity can be captured by semi-classical calculations \cite{anderson1972}, other properties must be accounted for by a fully quantum mechanical picture. Our lattice model  focuses on the formation and the heat capacity of the TLS, which can well be described semi-classically. Quantum properties, similar to molecular vibrations, cannot be studied with classical lattice model and are beyond the scope of this work. 
Despite these limitations, we will show that the TLSs that naturally emerge from the DPLM at low $T$ are fully consistent with and hence support the standard TLS picture, despite some technical differences to be explained below.


The standard TLS model is also widely referred to as the tunneling two-level (TTL) model. 
TTL model decribes the two-level system with the following Hamiltonian:
\begin{equation}
H=
\begin{pmatrix}
 E_1+\hbar \omega_1 & \hbar \omega_0\exp(-\frac{\delta}{2}(\frac{2mV}{\hbar^2})^\frac{1}{2}) \\ 
 \hbar \omega_0\exp(-\frac{\delta}{2}(\frac{2mV}{\hbar^2})^\frac{1}{2}) & E_2+\hbar \omega_2
\end{pmatrix}
\label{TLSH}
\end{equation}
where $E1,E2$ are energy of two quasi-stable configurations, $\omega_0,\omega_1,\omega_2$ are separately inter-well hopping frequency and intra-well oscillation frequency on site 1 and two. $V$ and $\delta$ are used to denote the energy barrier height and width. In their original paper \cite{anderson1972}, Anderson et al already pointed out that the relevant TLS should have large energy barrier so that resonant tunneling does not occur (V is large enough that off-diagonal elements are negligible), but transitions can occur by processes such as phonon-assisted tunneling. 
These transitions are thus incoherent processes consistent with a semi-classical description. The DPLM is basically a classical model. Particle hops at low $T$ should thus be interpreted in the semi-classical sense. The hopping rate $w$ in \eq{w} is then taken as a rough approximation for phonon-assisted tunneling processes. More accurate forms of particle hopping rates however should not alter our results qualitatively.

Our expression of $C_v$ from \eq{Cv} with $c_1$ from \eq{c} is equivalent to that in \Ref{anderson1972}, after neglecting $c_3T^3$ and identifying $z\phi_v \tilde P(\abs{\Delta E})$ with, in our notation, the density $n(\abs{\Delta E})$ of TLS per particle per unit energy in \Ref{anderson1972}  at $\abs{\Delta E}=0$, i.e.
\begin{equation}
  \label{nDE}
  n(0) = z \phi_v \tilde P(0).
\end{equation}

Moreover, the standard model assumes random particle hopping barriers uncorrelated to the TLS energy split $\abs{\Delta E}$ \cite{anderson1972}. Particles happen to have surmountable barriers constitute the TLS. In the DPLM, whether a particle can hop is also random, but the randomness primarily comes from whether it is a neighbor of a void. If a particle is next to a void so that a hop is allowed, the barrier then depends solely on the energy difference $\Delta E$ according to simple Metropolis rule in \eq{w} without further randomness. Nevertheless, the DPLM can be generalized to have additional randomness in the barriers, which should not alter our results qualitatively.

\subsection{Real space structure of TLS}
Despite decades of study, what constitutes the TLS is still controversial  \cite{muller2019}. 
The movement of rigid molecular groups suggested in the original paper of Anderson \textit{et al} \cite{anderson1972} is still the leading contender. Our picture basically follows this view.
A particle in the DPLM represents an atom or a rigid molecular group, while a void represents a quasivoid consisting of coupled free-volume fragments of a combined size comparable to that of a particle  \cite{yip2020}. Moreover, a TLS transition is identified with a microstring particle hopping motion, in which a short chain of particles displace one another synchronously \cite{glotzer2003}. They have been suggested as elementary motions in glasses \cite{chandler2011,lam2017}, a notion supported by colloidal experiments at high density \cite{yip2020}.
At present, the DPLM only directly simulates microstrings of unit length. 
Noting their strong back-and-forth nature
 as quantified by a high particle return probability $P_{ret}$, we have suggested that reversed microstrings are responsible for $\beta$ relaxations while only the non-reversed ones, which become increasingly few as $T$ decreases, lead to structural relaxations \cite{lam2017}. In this work, we further establish that as the void return probability $Q_{ret}$ approaches 1 at very low $T$, these microstrings constitute TLS transitions as well. 
These provide a simple unified view for these seemingly diverse processes of glasses.

\section{Conclusion}
\label{sec:conclusion}
To conclude, we have shown that the specific heat of the DPLM follows $C_v\propto T$ at very low $T$ in agreement with experiments. By closely monitoring the motions of particles and voids, we observe formation of TLS as random particle interactions induce  strong localization of voids to within two lattice sites. System relaxation is then limited to TLS transitions. For ultrastable glasses with a very low fictive temperature, the TLS density and thermal properties can be analytically calculated. For less stable glasses with fictive temperature close to the glass transition temperature, TLS emerge similarly at low $T$ after local relaxation subsides.

\begin{acknowledgments}
We thank the support of 
National Natural Science Foundation of China (Grants 11974297 and 11774022).
\end{acknowledgments}

\appendix

\section{Details of analytic calculation of specific heat capacity of TLS}
\label{sec:ana_cal}
We now provide further details on the calculation of the specific heat capacity in the DPLM. 
Consider a TLS with its initial state labeled 1 and the other state labeled  2. Denote the system energy at these two states by $E_1$ and $E_2$ so that $\Delta E = E_2 - E_1$. 
The relaxation rate $w_{TLS}$ of the TLS equals the sum of the forward and backward transition rates of the TLS, i.e. $w_{TLS} =w_{1\rightarrow 2} + w_{2\rightarrow 1}$, implying $w_{TLS} = w(\Delta E) + w(-\Delta E)$. In our simulations, we adopt particle hopping rates in the Metropolis form, i.e.
\begin{equation}
w(\Delta E) = 
\begin{cases}
w_0 e^{- \Delta E / k_BT} & \text{for} ~ \Delta E>0, \\
w_0  & \text{for} ~ \Delta E\leq 0. \label{si:w}
\end{cases}
\end{equation}
We thus get
\begin{equation}
  \label{si:wTLS}
  w_{TLS} = w_0 \roundbk{1 + e^{-\abs{\Delta E}/k_BT}} \ge w_0.
\end{equation}
All TLS in the DPLM thus relax fast and this explains their equilibrium nature even at very low $T$ as numerically demonstrated in \fig{voiddisp}(b).

Since TLS are at equilibrium, its average energy $\epsilon_{TLS}$ can be calculated using the Boltzmann distribution and we get
\begin{eqnarray}
\epsilon_{TLS}&=&\frac{E_1e^{- E_1/{k_BT}}+E_2e^{-E_2 /{k_BT}}}{e^{- E_1/{k_BT}}+e^{-E_2 /{k_BT}}}.
\end{eqnarray}
The heat capacity  
$\f(\Delta E) =
 {d \epsilon_{TLS}}/{dT}$ of a TLS is then given by
\begin{equation}
\f(\Delta E)  =
  \frac{1}{4k_BT^2} 
  \Delta E^2 \text{sech}^2\left(\frac{\Delta E}{2k_BT}\right) .
\end{equation}

Consider $T\ll T_g$ in which voids admit few energetically possible hopping pathways due to the strong localization.
Assume also a small void density $\phi_v$ so that voids are isolated. The initial equilibrium position of a void is associated with state 1 of a possible TLS. There is a  probability $q$ that the void can hop to a given nearest neighboring occupied site with an energy cost smaller than $\Delta E_{max}$, realizing a TLS transition to state 2. Taking into account all possible TLS, the specific heat capacity $C_v$, defined as heat capacity per particle, is 
\begin{equation}
C_v = {z \phi_{v} }{} \int^\DEmax_{-\infty} d\Delta E ~ {P}(\Delta E) \f(\Delta E), \label{si:Cv2}
\end{equation}
where $z=4$ is the lattice coordination number and $P(\Delta E)$ is the probability distribution of $\Delta E$.
The void density $\phi_v$ has been assumed a constant independent of $T$, as is assumed in our simulations for simplicity.

Only TLS with an energy split $\DEabs$ within a few $k_BT$ can contribute significantly to the heat capacity. This is reflected in the function $\f(\Delta E)$, which is sharply peaked at $\Delta E \approx \pm 2.35 k_BT$. 
The upper integration limit in \eq{si:Cv2} can thus be approximated by infinity. Noting also that  $\f(\Delta E)$ is an even function of $\Delta E$, \eq{si:Cv2} gives
\begin{equation}
C_v = z \phi_{v}  \int^\infty_{0} d \DEabs ~ \tilde{P}(\DEabs) \f(\DEabs),
\label{si:Cvtheory2}
\end{equation}
where the TLS energy split $\DEabs$ has a distribution
\begin{equation}
\tilde{P}(\DEabs) = P(\Delta E) + P(-\Delta E).
\label{si:Ptilde2}
\end{equation}
The distribution $P(\Delta E)$ is a smooth function provided the interaction distribution $g(V)$ is sufficiently smooth, which should hold true in realistic systems. We expand $P(\Delta E)$ about $\Delta E =0$ and write
\begin{equation}
  \label{si:series}
  P(\Delta E) = P(0) + \Delta E~ P'(0)  + \frac12 \Delta E^2 P''(0) + \dots.   ~~
\end{equation}
Then, \eq{si:Ptilde2} becomes
\begin{equation}
\tilde{P}(\DEabs) = 2P(0) + \Delta E^2 P''(0) + \dots.  
\label{si:Ptilde3}
\end{equation}
All odd-power terms vanish exactly as 
$\tilde{P}(\DEabs)$ is an even function of $\DEabs$. 
Substituting \eq{si:Ptilde3} into \eq{si:Cvtheory2} and neglecting higher order terms, we get
\begin{equation}
  \label{si:Cv}
  C_v = c_1 T + c_3 T^3
\end{equation}
where
\begin{equation}
c_1 = \frac{z}2 I_1\phi_v k^2_B P(0), ~~~~~
c_3 = \frac{z}4  I_3 \phi_v k^4_B P^{''}(0). 
\label{si:c0}
\end{equation}
We have defined
\begin{equation}
I_n = \int^\infty_0 dx ~ x^{n+1} \text{sech}^2(x/2)
\end{equation}
so that $I_1=2\pi^2/3$ and $I_3=14\pi^4/15$. These give
\begin{equation}
c_1 = \frac{\pi^2z\phi_v k^2_B P(0)}{3}, ~~~~~
c_3 = \frac{7\pi^4z\phi_v k^4_B P^{''}(0)}{30}. 
\label{si:c}
\end{equation}
An interesting observation is that all even terms, e.g. $c_2 T^2$, vanish exactly, which follows directly from the vanishing of all odd terms in $\tilde{P}(\DEabs)$ from \eq{si:Ptilde3}.

We now further assume an ultrastable system equilibrated at an initial temperature $\Ti \ll T_g$. Exact equilibrium properties of the DPLM \cite{zhang2017} then allow an exact evaluation of $C_v$.
At equilibrium temperature $\Ti$, the interaction $V_{s_is_j}$ between particles occupying sites $i$ and $j$ follows a distribution 
$p_{eq}(V)$, which is simply the Boltzmann distribution \cite{zhang2017,lulli2020}
 \begin{equation}
 p_{eq}(V)=\frac{g(V)e^{ -{V}/{k_B \Ti} }}{\int^{\infty}_{-\infty}g(V)e^{ -{V}/{k_B\Ti} }\mathrm{d}V}
 \end{equation}
Starting from the initial state 1 of the TLS,  a  given hop attempt to attain state 2 involves an energy change $\Delta E$ of the system given by
\cite{lam2018}
\begin{equation}
\Delta E = \sum_{\gamma=1}^{z-1} \roundbk{V'_\gamma - V_\gamma}\\
\label{si:DE}
\end{equation} 
where $V_\gamma$ denotes $z-1$ initial interactions  to be broken and $V'_\gamma$ denotes $z-1$ new interactions to be formed. Here, $V_\gamma$  follows the a posteriori distribution $p_{eq}(V)$ because they are realized in the initial equilibrium configuration. In contrast $V'_\gamma$ follows the a priori distribution $g(V)$ because without stipulating that the hop attempt must be successful, any new interactions are equally likely. 

\begin{figure}[tpb]
\centering
\includegraphics[width=0.45\textwidth]{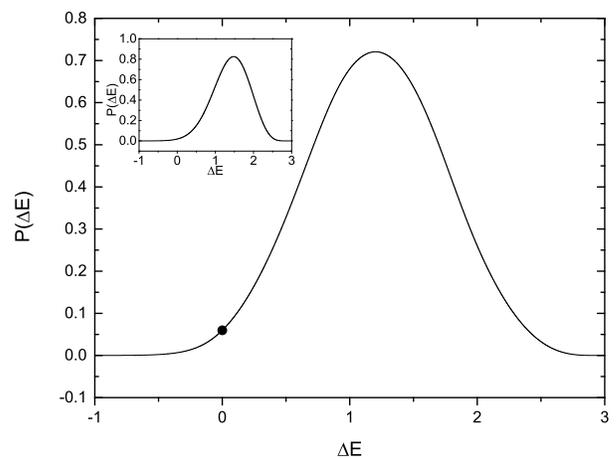}
\caption{ Probability distribution $P(\Delta E)$ of energy change $\Delta E$ due to level switching of a TLS. The system is equilibrium at temperature $\Ti=0.1$. Despite using a uniform distribution $g(V)$ of particle interactions, $P(\Delta E)$ is $not$ flat. The low-$T$ heat capacity coefficient $c_1$ depends only on $P(0)$ (black dot). Inset: $P(\Delta E)$ against $\Delta E$ for an alternative non-uniform $g(V)=2V$ for $V\in [0,1]$, leading to a qualitatively similar $P(\Delta E)$. 
}
\label{PDE} 
\end{figure}

Note that $C_v$ in \eq{si:Cvtheory2} depends on the coordination number $z$ not only explicitly but also implicitly via $\tilde{P}(\DEabs)$. Moreover, $z$ in turn depends on the lattice type and more generally on the system dimension.
We now take $z=4$ for the square lattice adopted in this work. \eq{si:DE} states that $\Delta E$ is a sum of six random variables and its distribution thus follows the convolution form
 \begin{equation}
 P(\Delta E)=
 ({g\circ g\circ g}\circ {P_{eq}^{-}\circ P_{eq}^{-}\circ P_{eq}^{-}})(\Delta E)
\label{si:PDE}
\end{equation}
where 
$P_{eq}^{-}(V)=P_{eq}(\Delta V-V)$, which is non-zero for $V \in [0, \Delta V]$. 
In general,  $P(\Delta E)$ can be evaluated numerically using \eq{si:PDE} for any $g(V)$. \fig{PDE} shows the numerical result of $P(\Delta E)$.
Note that $P(\Delta E)$ is not a flat distribution as often assumed for simplicity \cite{phillips1987}, despite a uniform interaction distribution $g(V)$ being used. According to \eq{si:c}, the low-$T$ heat capacity depends only on $P(0)$ and $P''(0)$ and other details of $P(\Delta E)$ is irrelevant. For comparison, the inset in \fig{PDE} shows $P(\Delta E)$ for a different $g(V)=2V$ for $V\in [0,1]$. We observe a qualitatively similar $P(\Delta E)$, which will also lead to qualitatively similar heat capacity properties predictable using \eq{si:c}. 

\begin{figure}[pbt]
\centering
\includegraphics[width=0.45\textwidth]{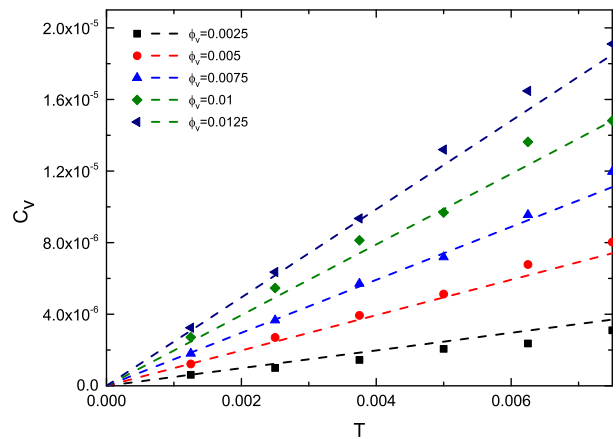}
\caption{Specific heat capacity $C_v$ against temperature $T$ for initial temperature $\Ti=0.05$ and various void density  $\phi_v \le 0.0125$. Dashed lines show theoretical values from \eqs{si:Cv}{si:c1c2}.  
}
\label{CvTphiv}
\end{figure}
\begin{figure}[tpb]
\centering
\includegraphics[width=0.45\textwidth]{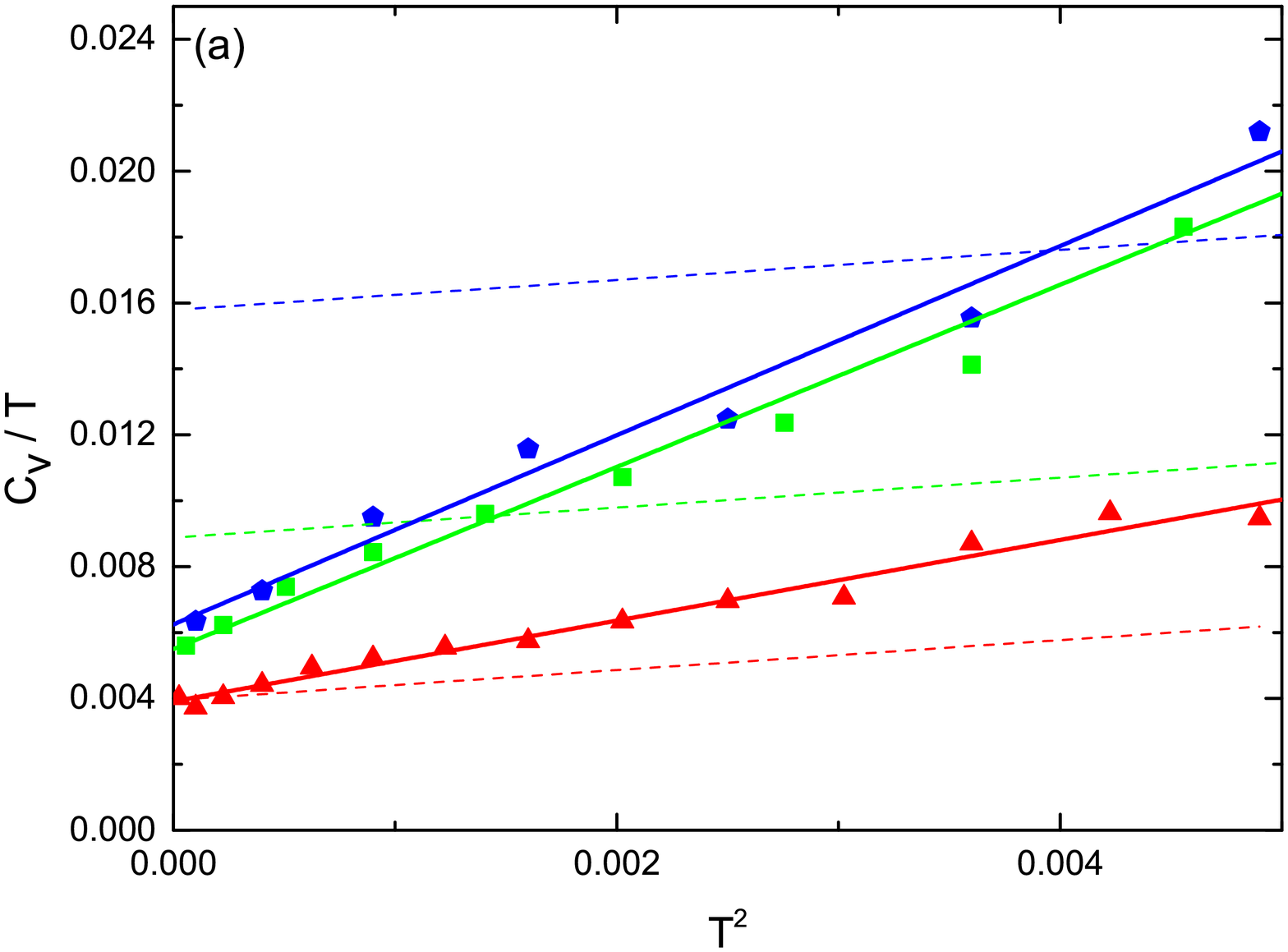}
\includegraphics[width=0.45\textwidth]{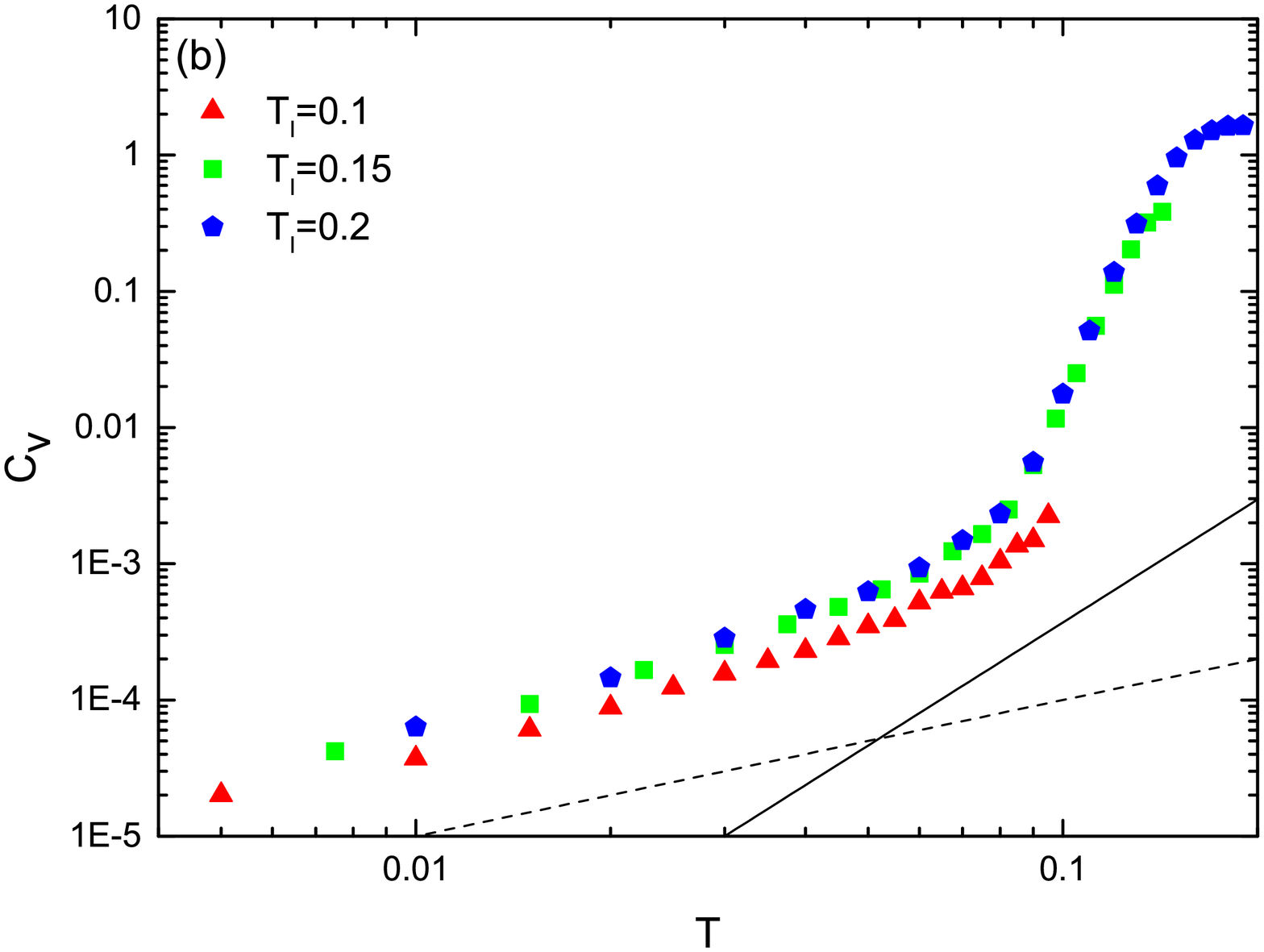}
\caption{ (a) Plot of $C_v/T$ against $T^2$ at low $T$ during cooling from a high initial temperature $\Ti$ not well below $T_g$. It satisfies  \eq{si:Cv} as shown by the good fits to linear relations (solid lines). However, as $\Ti$ increases, results increasingly deviate from the coefficients $c_1$ and $c_3$ predicted in \eq{si:c1c2} (dotted lines) because of local relaxation into non-equilibrium configurations. 
(b) $C_v$ against $T$ in log-log scales from the same simulations as in (a) plotted over the full range of $T$. As $T$ decreases, $C_v$ drops from values up to 1.6 $k_B$ , with $k_B=1$, to small values, after which it follows \eq{si:Cv}. The dashed and solid lines with slopes 1 and 3 are guides to the eye. 
}
\label{highTi} 
\end{figure}
\begin{figure*}[tb]
\centering
\includegraphics[width=0.4\textwidth]{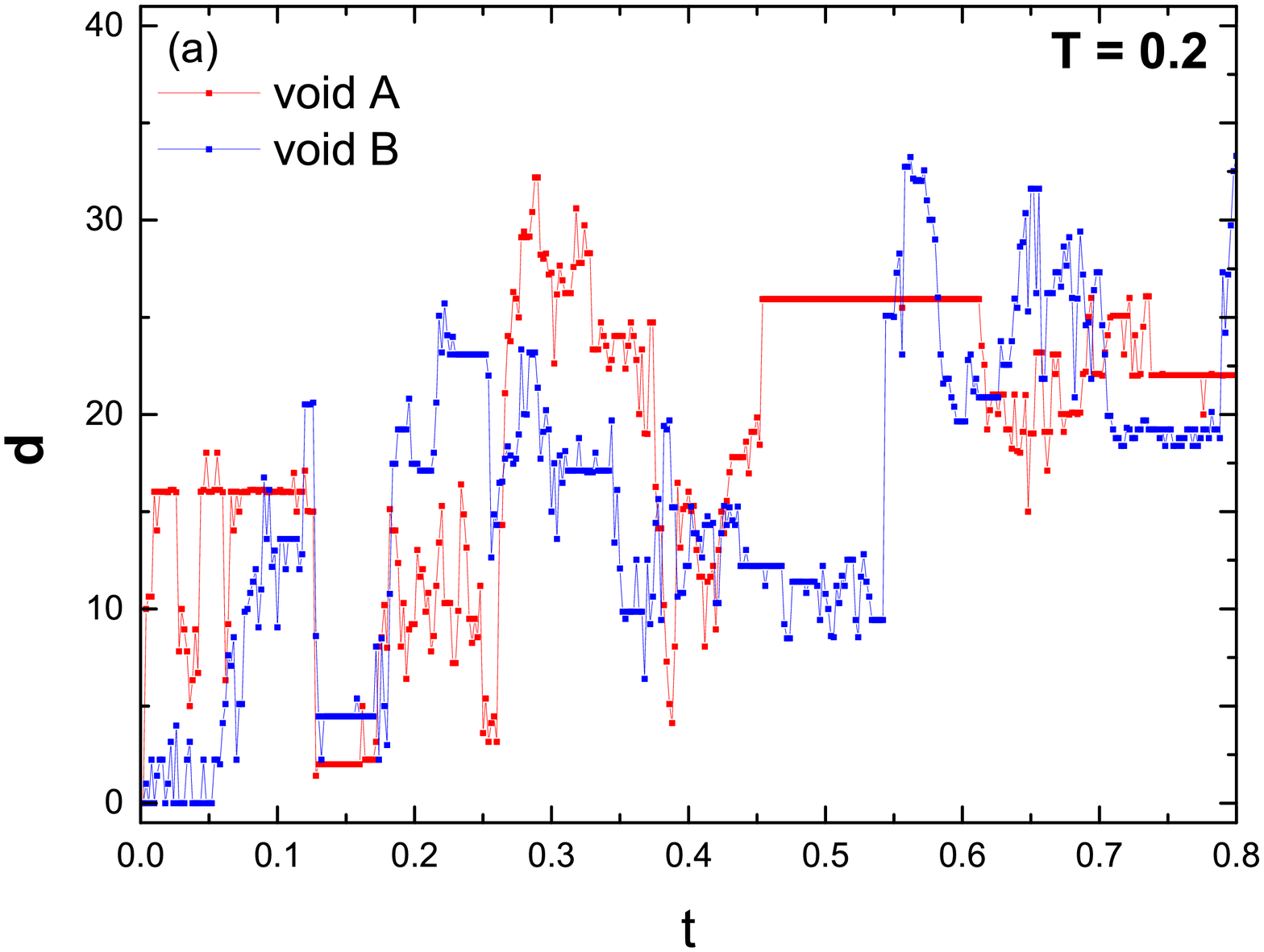}
\includegraphics[width=0.4\textwidth]{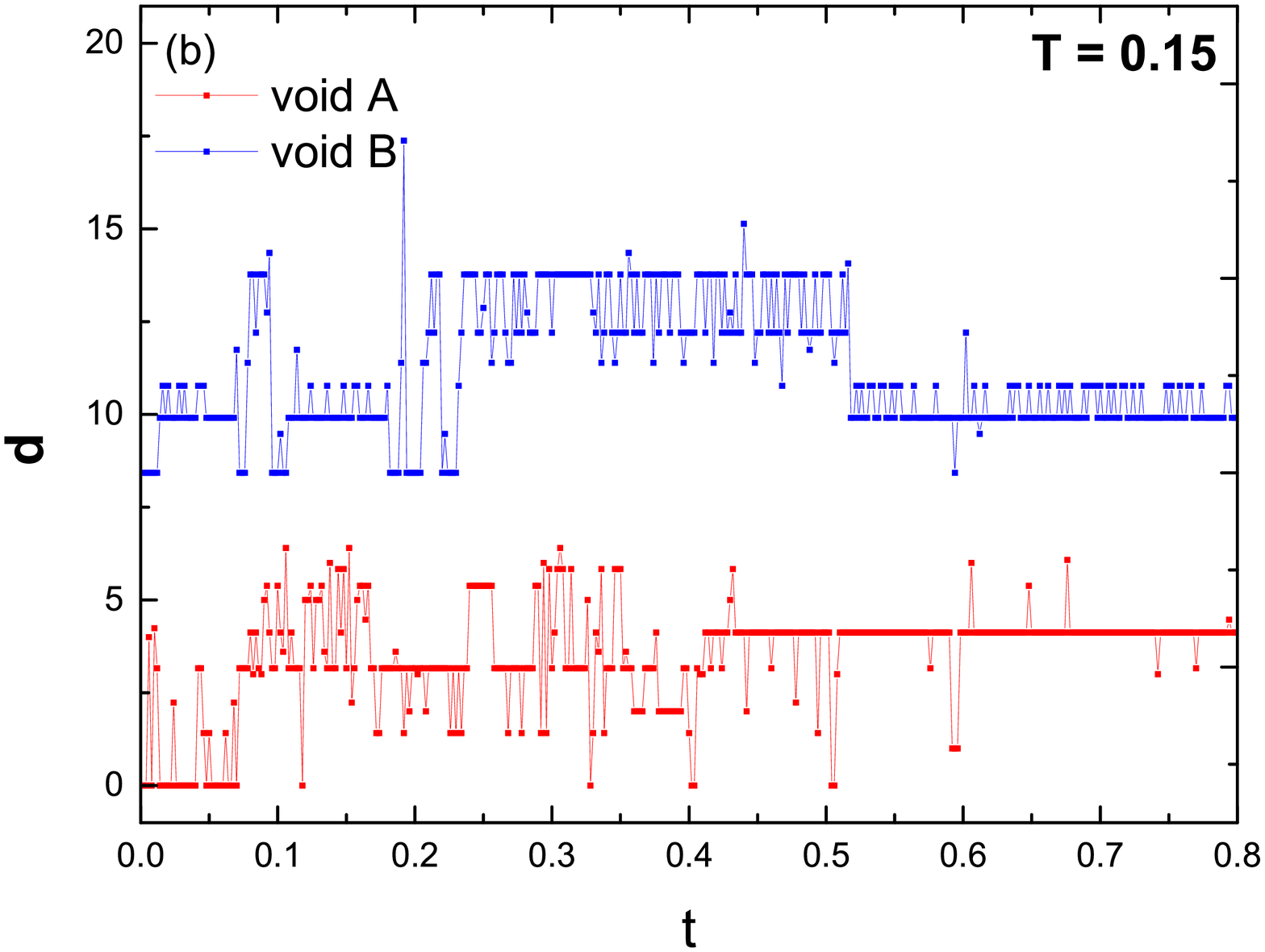}
\includegraphics[width=0.4\textwidth]{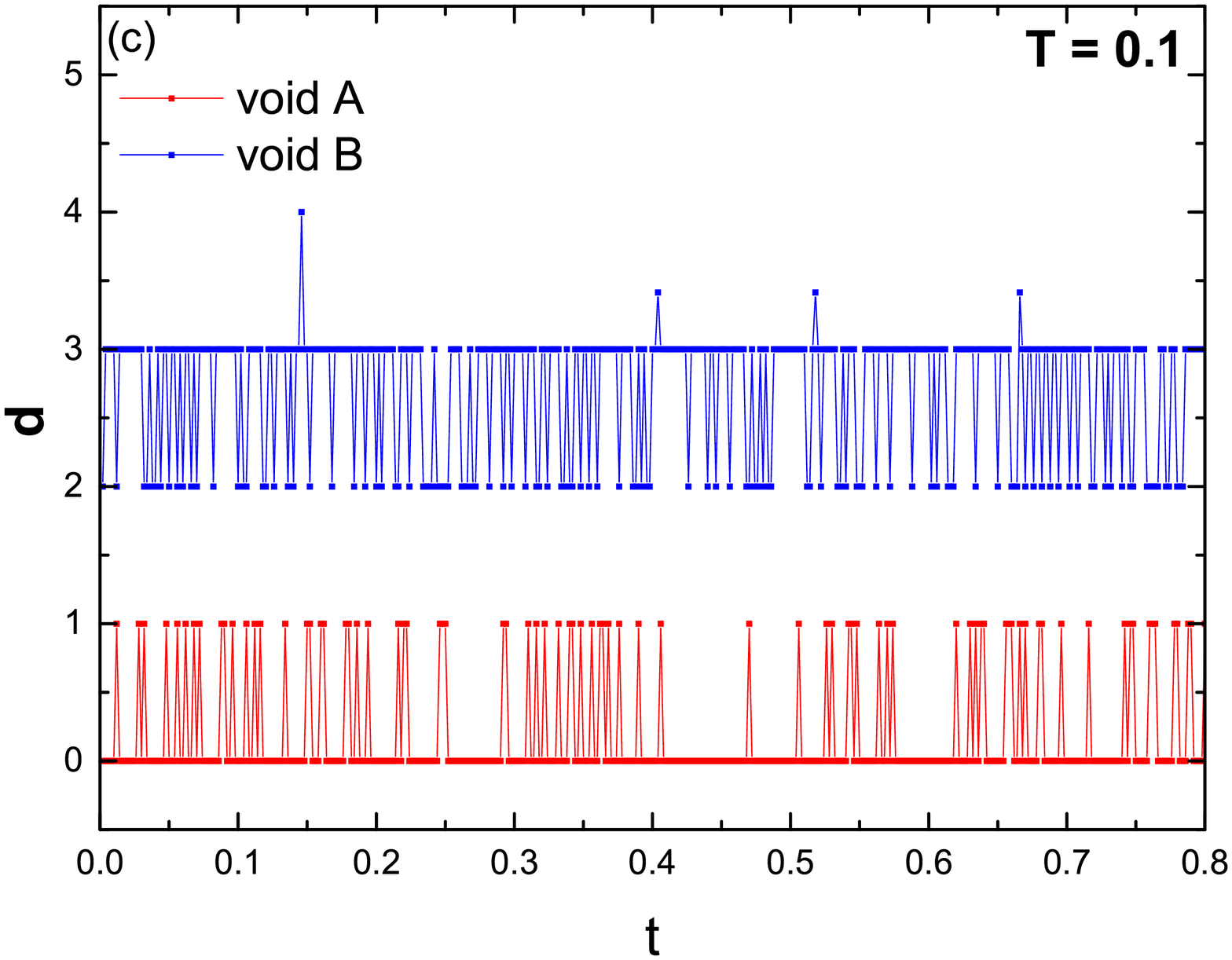}
\includegraphics[width=0.4\textwidth]{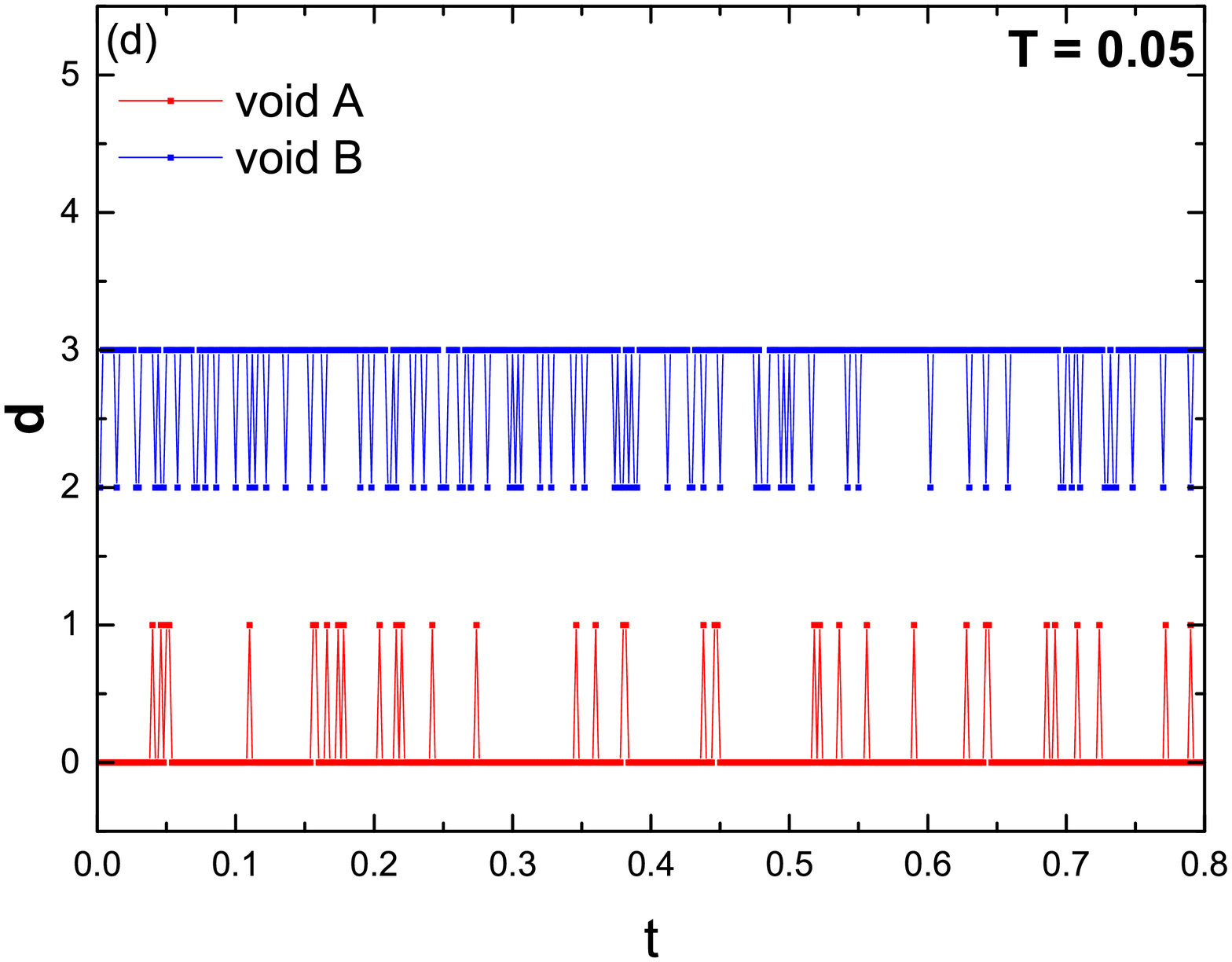}
\caption{Plot of void displacement $d$ against time $t$ of two TLS in Fig.~\href{twolevel-main}{2} in the main text for $T=$0.2 (a), 0.15 (b), 0.1 (c)  and 0.05 (d). Results for TLS B are shifted upward for clarity. 
}
\label{si:voiddisp}
\end{figure*}

In particular, consider the interaction distribution $g(V)$ uniform in $[0, \Delta V]$ adopted in our main simulations. Analytic calculation is possible.
Performing simple algebra in the Laplace tranformed space, 
 \eq{si:PDE} becomes
 \begin{equation}
  P(\Delta E)= \mathscr{L}^{-1}[(\mathscr{L}[g])^{3} (\mathscr{L}[P_{eq}^{-}])^{3})](\Delta E + 3\Delta V).
 \end{equation}
where $\mathscr{L}$ denotes the Laplace transform. 
The relevant derivatives at $\Delta E=0$ are found to be, after some algebra, 
 \begin{equation}
P(0) = \frac{6k_B^2\Ti^2}{\Delta V^3}, ~~~~~ P^{''}(0) = \frac{1}{\Delta V^3}.\label{si:P}
 \end{equation}
Substituting into \eq{si:c}, we get
\begin{equation}
c_1 = \frac{8\pi^2 \phi_v k^4_B \Ti^2}{ \Delta V^3},  ~~~~~~ c_3 = \frac{14\pi^4\phi_vk^4_B}{15\Delta V^3}.
\label{si:c1c2}
\end{equation}
Note that $c_1$ and $c_3$ from \eq{si:c1c2} are exact in the limit $T \le T_I \ll T_g$ corresponding to ultrastable glasses, in which the only relaxation modes are TLS relaxations.
They are accurately verified by DPLM simulations under these conditions as shown in \fig{CvT}.

Generalization to other lattice coordination number $z$ is straight-forward. For example, for  $z=6$ appropriate for a triangular lattice in 2D or a cubic lattice in 3D, we get
 \begin{equation}
 P(0) = \frac{70k_B^4T_I^4}{\Delta V^5}, ~~~~ P^{''}(0) = \frac{15k_B^2T_I^2}{\Delta V^5}\label{si:P2}
 \end{equation}
and hence 
 \begin{equation}
 c_1=\frac{140\pi^2 \phi_v k^6_B\Ti^4}{\Delta V^5},
~~~ c_3 =\frac{21\pi^4\phi_vk^6_B\Ti^2}{\Delta V^5}. \label{si:R2}
 \end{equation}
For more general forms of $g(V)$, the Laplace transform may become intractable analytically but $c_1$ and $c_3$ can be readily solved accurately by performing the convolution numerically.

\section{Supplemental simulation results}

Our main simulations have been performed using a void density $\phi_v = 0.005$. We have also performed simulations using a wider range of $\phi_v$ and results on $C_v$ are plotted in \fig{CvTphiv}. Good agreement with \eqs{si:Cv}{si:c1c2} is observed.
In particular, \eq{si:c1c2} implies that $c_1 \propto \phi_v$ which is well verified here. It shows that the TLS in the system are isolated and independent of each other at  small $\phi_v$. 

\begin{figure}[tpb]
\centering
\includegraphics[width=0.45\textwidth]{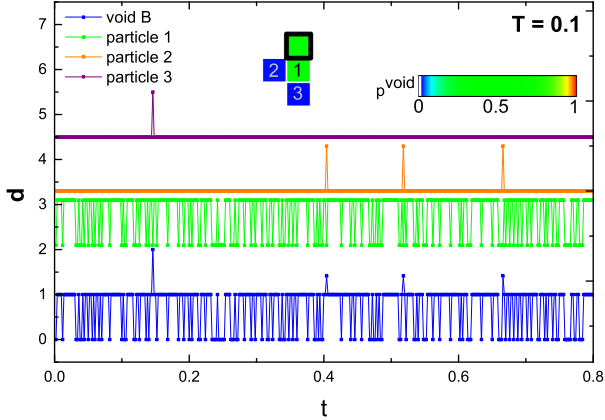}
\caption{Plot of displacement $d$ against time $t$ for a four-level system at $T=0.1$ resulting from the motion of void B and three related particles. 
Curves for particles are shifted upward for clarity. Inset: Void occupation probability $\pvoid$ in the region containing the four-level system. A black square marks the position of the void at $t=0$. Initial particle positions are labeled as 1, 2, and 3. 
}
\label{voidandparticle}
\end{figure}

We have focused on $T \le \Ti \ll T_g$ corresponding to ultrastable glasses, for which analytical expressions are obtained. 
We now explain additional simulations on less  stable glasses with a higher initial temperature $\Ti$. 
\Fig{highTi}(a) plots $C_V/T$ against $T^2$ at low $T$. Results are consistent with \eq{si:Cv}, although $c_1$ and $c_3$, i.e. the y-intercept and slope, deviate from the theoretical values in \eq{si:c1c2}. The discrepancies increase with $\Ti$ because the initial temperature $\Ti$ can no longer be taken as the fictive temperature at low $T$ due to  significant relaxations. 
To illustrate the full picture, \fig{highTi}(b) plots $C_v$ against $T$ in a log-log scale from the same simulations for the entire temperature range. Consider $\Ti = 0.15$ or 0.2 simulating the formation of glasses by cooling from the liquid phase. At high $T$, $C_v$ is of the order of $k_B$, where $k_B=1$. This is consistent with typical experimental values of excess entropy of glasses over their crystalline counterparts \cite{angell2000}. As $T$ decreases, $C_v$ drops by a few orders of magnitude and eventually follows the temperature dependence in \eq{si:Cv}. Note that the curves for $\Ti = 0.15$ and 0.2 in \fig{highTi}(a) and (b) nearly coincide  at $T\le 0.15$. This is because the systems remain close to equilibrium during cooling at $T\agt T_g \simeq 0.15$ so that the thermal history above $T_g$ is irrelevant.

\fig{voiddisp}(a) in the main text plots the void displacement versus time of two TLS at $T=0.05$, revealing their bistable nature. To provide the full picture, they are reproduced in \fig{si:voiddisp}, which also shows similar displacement-time graphs of these two voids at a wide range of $T$.  We observe that at $T = 0.2 \gg T_g$, the voids are mobile and the displacements resemble those of simple random walks, indicating the liquid phase.  At $T=0.15 \simeq T_g$, localization of the voids during the displayed period is clear. At $T=0.1 \ll T_g$, the voids are much more tightly localized. The system is deep in the glass phase. One void already forms a TLS. The other leads to a four-level system, although the two excited levels carry much less probabilistic weights. At $T=0.05$, both TLS have emerged from the strong localization, without detectable transition to higher levels. 

    In the main text, we have argued that a TLS requires a void return probability $Q_{ret}=1$, while a particle return probability $P_{ret}=1$ is a necessary but an insufficient condition. To explain it further, \fig{voidandparticle} shows the displacement-time graph of a void exhibiting a four-level system. A transition between the two lower energy  levels with $d=0$ and 1 involves the hop of a particle, the displacement of which is also shown (green). Excitations to two other levels with $d=\sqrt 2$ and $2$ in contrast involve the hop of two other particles, with their displacements also shown (orange and purple). From their displacement-time graphs, all three particles exhibit bistability and contribute to a unit particle return probability $P_{ret}$. However, they do not form three non-interacting TLS, as the first particle must be at the $d=1$ state before one of the other two particles can hop. These constraints are easily understood from the spatial profile of the possible positions of the void (see inset in \fig{voidandparticle}). Therefore, this void together with the three particles form a four-level system, rather than three independent TLS. 

    Note that in the displacement-time graphs of voids and particles discussed above, $d$ alone does not perfectly resolve all possible levels. For example, $d=1$ can result from one of any four possible nearest neighboring hops of the void on a square lattice. Thus, we have also examined real space images as well as $x$ and $y$ components of the void displacement. All examples of TLS described by these plots indeed exhibit bistability. 

~~

\end{document}